\documentclass{PoS}
\usepackage{siunitx}
\usepackage{url}
\usepackage{textcomp}
\usepackage{graphicx,subcaption}
\usepackage[percent]{overpic}

\DeclareSIUnit\permille{\text{\textperthousand}}

\title{Silicon Vertex \& Tracking Detectors for the Compact Linear Collider}

\ShortTitle{Silicon Vertex \& Tracking Detectors for CLIC}

\author{\speaker{Simon Spannagel} {\normalfont on behalf of the CLICdp Collaboration}\\
        CERN\\
        E-mail: \email{simon.spannagel@cern.ch}}

\abstract{
CLIC is a proposed linear $e^+e^-$ collider with center-of-mass energies of up to \SI{3}{TeV}.
Its main objectives are precise top quark, Higgs boson and Beyond Standard Model physics.
In addition to spatial resolutions of a few micrometers and a very low material budget, the vertex and tracking detectors also require timing capabilities with a precision of a few nanoseconds to allow suppression of beam-induced background particles.
Different technologies using hybrid silicon detectors are explored for the vertex detectors, such as dedicated \SI{65}{\nano \meter} readout ASICs, small-pitch sensors as well as bonding using anisotropic conductive films.
Monolithic sensors are the current choice for the tracking detector, and a prototype using a \SI{180}{\nano \meter} high-resistivity CMOS process has been designed and produced, and is currently under evaluation.
Different designs using a silicon-on-insulator process are under investigation for both vertex and tracking detector.
All prototypes are tested in laboratory and beam tests, and newly developed simulation tools combining Geant4 and TCAD are used to assess and optimize their performance.
This contribution gives an overview of the R\&D program for the CLIC vertex and tracking detectors, highlighting new results from the prototypes.
}

\FullConference{The 28th International Workshop on Vertex Detectors - Vertex2019\\
		13-18 October, 2019\\
		Lopud, Croatia}

\begin{document}

\section{Introduction}

The Compact Linear Collider (CLIC)~\cite{clic,clic-baseline,clic-summary} is a proposed $e^+e^-$ linear collider at CERN for the era beyond HL-LHC.
Its main objectives are precise top quark, Higgs boson and Beyond Standard Model physics.
The construction is planned in three energy stages of \SI{380}{GeV}, \SI{1.5}{TeV} and \SI{3}{TeV}.
The accelerator employs a novel and unique two-beam acceleration method:
A high-current, low-energy drive beam is used to accelerate the high-energy main beam by transferring energy from one beam to the other in normal-conducting so-called two-beam modules.
Using this technique, acceleration gradients of more than \SI{100}{MV/m} can be achieved.
A detector to be operated at the collision point of CLIC is being developed by the CLIC Detector \& Physics (CLICdp) collaboration, and an advanced detector model has been presented~\cite{dettech-yr,clicdet}.

This contribution focuses on the research and development undertaken for the vertex and tracking detector technologies of the experiment.
The document is structured as follows.
Section~\ref{sec:condition} describes the experimental conditions at the collision point with a focus on the parameters relevant for the vertexing and tracking system.
The currently envisaged layout of these detectors is described in Section~\ref{sec:layout}.
Subsequently, recent results from the investigation of hybrid (Section~\ref{sec:hybrid}) and monolithic (Section~\ref{sec:monolithic}) silicon detectors are presented.
Section~\ref{sec:tools} describes several tools developed as part of the CLIC vertex \& tracking detector R\&D and available to other users within the community.
Finally, Section~\ref{sec:summary} provides a summary and an outlook to further developments.

\section{Experimental Conditions at CLIC}
\label{sec:condition}

CLIC operates in bunch trains with a repetition rate of \SI{50}{Hz}.
At \SI{3}{TeV} there are 312~bunches within each train with a bunch-to-bunch separation of \SI{0.5}{ns}.
In order to reach high luminosities of $\mathcal{L} = \SI{5.9e34}{\per \centi \meter \squared \per \second}$, the beams are focused at the interaction point to a transverse size of about $\sigma_x = \SI{40}{nm}$ and $\sigma_y = \SI{1}{nm}$.
The bunches of these nanometer beams create so-called beamstrahlung by interacting with the strong fields of the oncoming bunch.
The resulting beam-induced background consists of $e^+e^-$ pairs as well as charged and neutral hadrons and is predominantly emitted along the beam axis.
The flux of background particles defines the minimum size of the outgoing beam pipe and the inner detector layers as well as the granularity of these detectors to keep the occupancy below \SI{3}{\percent} per bunch train.

The very short bunch spacing as well as the number of background particles from beamstrahlung drives the timing requirements for the detector systems.
For the vertex and tracking detectors, a single-hit time resolution of about \SI{5}{ns} is required in order to efficiently separate beam-induced background, present throughout the full bunch train, from the hard physics event.

Owing to the low duty cycle of the accelerator, the detector has to be active only \SI{156}{\nano \second} of the \SI{20}{ms} cycle.
This allows for reading out all data recorded during the bunch crossings without a trigger.
Furthermore, it enables power pulsing, i.e.\ switching detector components to a low-power state between bunch trains in order to reduce power consumption and therefore heat dissipation.

\section{Vertex \& Tracking Detectors at CLIC}
\label{sec:layout}

The layout of the vertex and tracking detectors are driven by the experimental conditions described above as well as the requirements arising from full-simulation physics studies.

\begin{figure}[tbp]
        \centering
        \begin{subfigure}[t]{0.48\textwidth}
                \begin{overpic}[width=\textwidth]{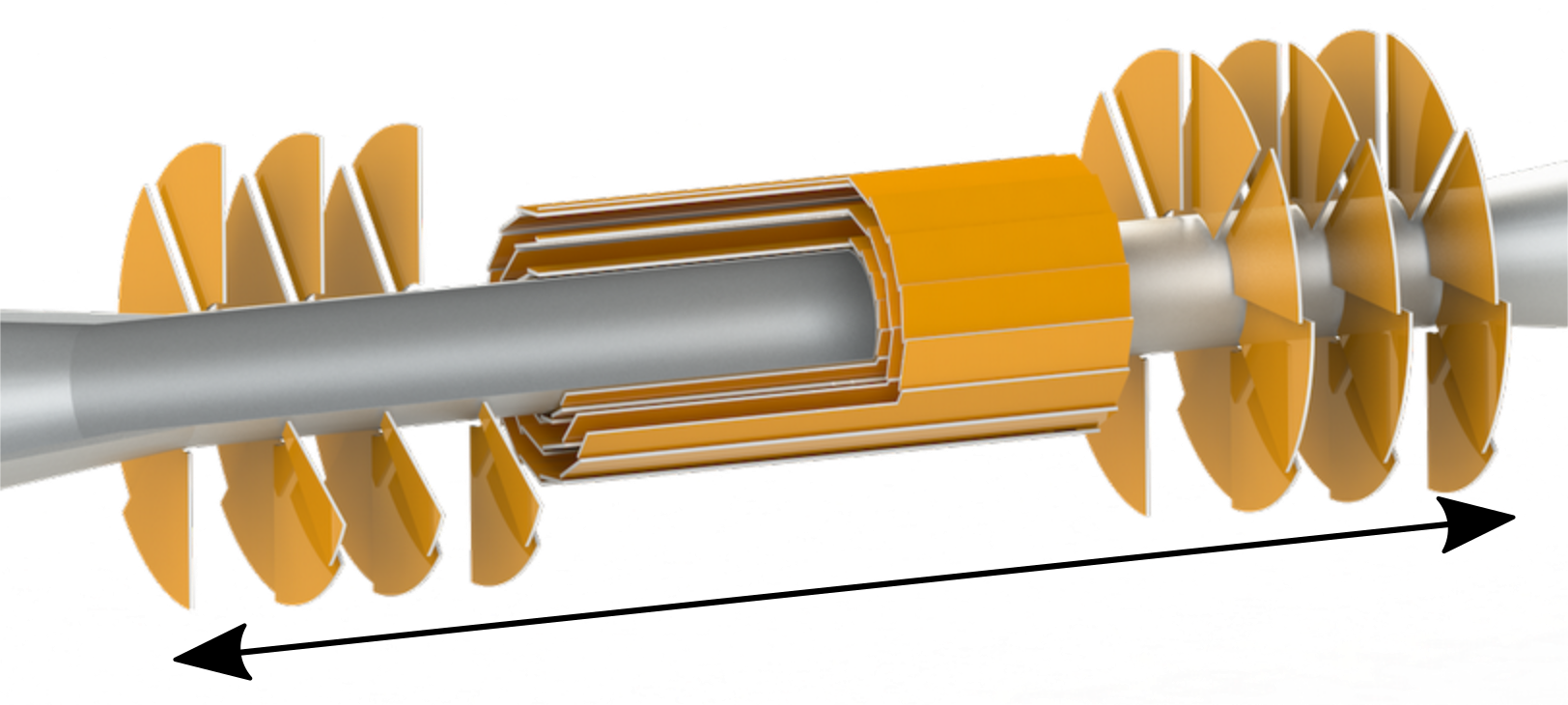}
                        \put (60,2) {\SI{600}{\milli \meter}}
                \end{overpic}
                \caption{Vertex detector with barrel and spiral end disks}
                \label{fig:clic:vertex}
        \end{subfigure}
        \begin{subfigure}[t]{0.48\textwidth}
                \begin{overpic}[width=\textwidth]{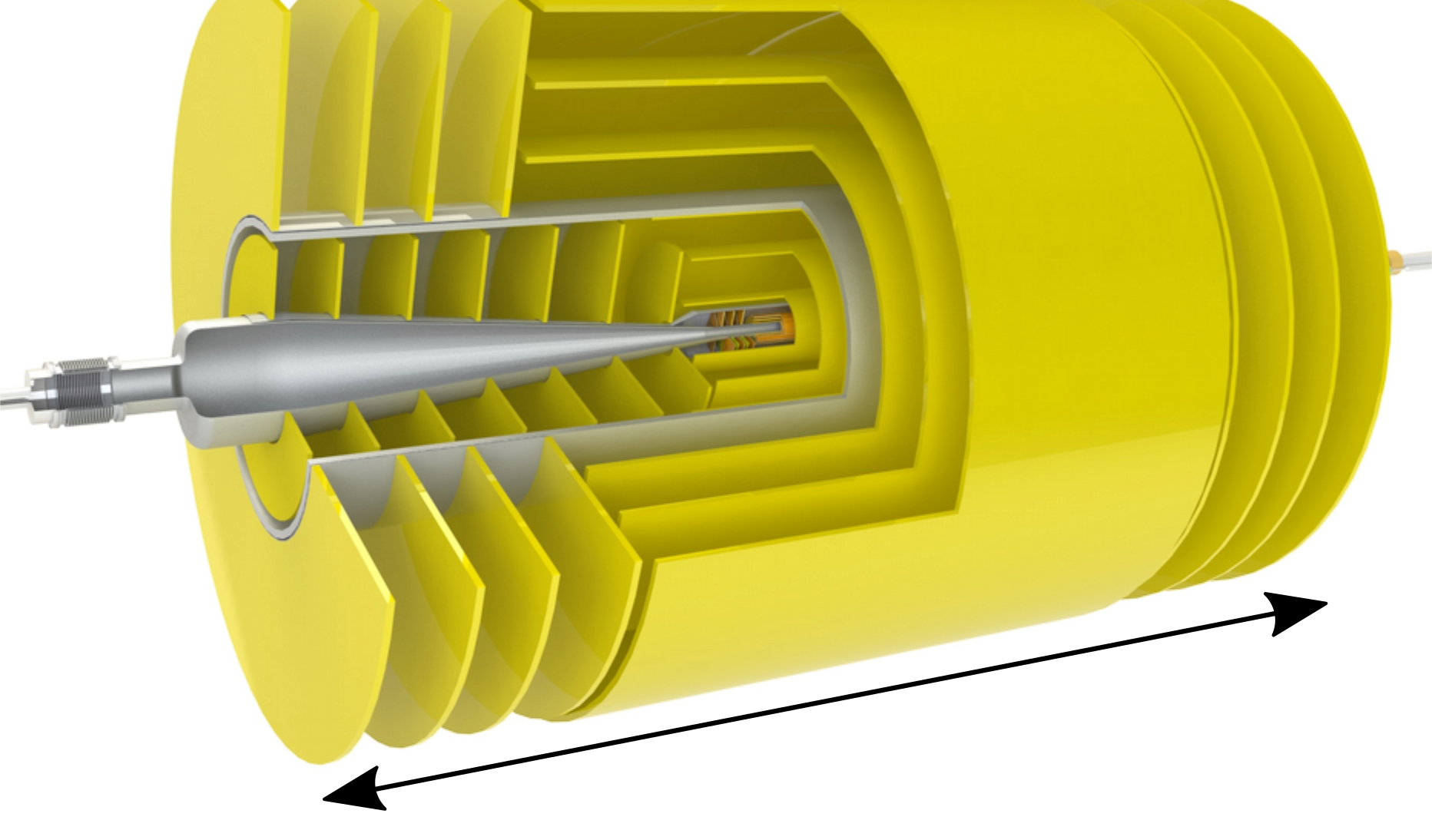}
                        \put (60,5) {\SI{4.4}{\meter}}
                \end{overpic}
                \caption{inner and outer tracking detector}
                \label{fig:clic:tracker}
        \end{subfigure}
        \caption{Designs for the CLIC vertex and tracking detectors as implemented in the CLICdet detector model. The drawings show the respective detectors as well as the beam pipe. From~\cite{clic-summary,dettech-yr}}
        \label{fig:clic}
\end{figure}

\subsection{Vertex Detector}

The design of the vertex detector is strongly driven by flavor tagging.
The goal is to build a detector which contains minimal scattering material of less than \SI{0.2}{\percent} $X_0$ per detection layer, combined with a high single-point resolution of $\sigma_{SP} \approx \SI{3}{\micro \meter}$.
To reach the ambitious material budget goal of the detector, the power consumption is limited to below \SI{50}{mW \per \centi \meter \squared} to allow for forced air-flow cooling~\cite{clicdp-note-2016-002}.
This limit is reached by means of power pulsing.
The detector front-ends are switched to a low power stand-by mode between the bunch trains as will be discussed in more detail in Section~\ref{sec:hybrid}.

The current prototype design comprises hybrid pixel detectors, arranged in three double layers in the barrel and three double disks at each end as shown in Figure~\ref{fig:clic:vertex}.
The inner barrel layer is constrained to a minimum of \SI{31}{\milli \meter} by the beam-induced backgrounds.
The disks are spirally shaped to direct the airflow through the detector for optimum cooling performance.
Each sensor layer consists of \SI{50}{\micro \meter} active silicon sensor plus \SI{50}{\micro \meter} of silicon for the readout ASIC, the pixels are foreseen to have a pitch of \SI{25 x 25}{\micro \meter} to cope with the expected background particle occupancies at \SI{3}{TeV} collisions.
This layout amounts to an active surface area of \SI{0.84}{\meter \squared}.

\subsection{Tracking Detector}

The figures of merit for the tracking detector are a good momentum resolution and a high tracking efficiency.
The required single-point resolution in the transverse plane has been estimated to $\sigma_{SP} \approx \SI{7}{\micro \meter}$ from full Monte Carlo simulations while the maximum allowed granularity in longitudinal direction of $1 - \SI{10}{\milli \meter}$ is determined by the occupancy arising from background particles~\cite{clicdp-note-2017-002}.
With a total length of \SI{4.4}{\meter}, the material budget of the detector is not only driven by the reduction of multiple scattering but also by the required rigidity of the structure holding the sensor modules.
Currently, monolithic silicon detectors with elongated pixels and a thickness of up to \SI{200}{\micro \meter} including electronics are under investigation as sensor technology.

The current design foresees leakless water cooling to room temperature and a lightweight support structure which amounts to a combined material budget of $1 - \SI{2}{\percent} X_0$ per detector layer.
The detector layout currently foreseen comprises six barrel layers and seven endcap layers with a total active area of approximately \SI{140}{\meter \squared} as indicated in Figure~\ref{fig:clic:tracker}.

\section{Hybrid Pixel Detectors}
\label{sec:hybrid}

Hybrid pixel detectors consist of two separate silicon layers, one being the sensor and the other the readout ASIC housing the front-end electronics.
This separation allows to use available and well-understood technologies for both wafers.
The sensor on one hand is produced in high-resistivity silicon with segmented electrodes, and high bias voltages can be applied to fully deplete the volume.
The CMOS readout chip on the other hand can be designed using mixed-mode CMOS circuits with small feature sizes to incorporate extensive functionality into the individual pixel cells.
In most cases, solder bumps act as interconnects between the two wafers, and bump bonding with small pixel pitches of down to \SI{25}{\micro \meter} has been achieved~\cite{dettech-yr}.

The downside of this approach is the requirement for bump bonding, which is a limiting factor for both the pixel pitch and the device thickness due to the bump size and bonding process.
Furthermore, bump bonding is a cost driving factor in manufacturing of hybrid pixel detector modules.

For the CLIC detector studies, several dedicated hybrid prototypes have been designed and tested in laboratory and test beam measurements in order to identify challenges and to qualify technologies for the requirements outlined above.

\subsection{The CLICpix2 Prototype}

The CLICpix2 prototype is the latest hybrid pixel detector ASIC designed to meet the CLIC vertex detector requirements~\cite{clicpix2}.
It is fabricated in a \SI{65}{\nano \meter} CMOS process and shares some design features with the ASICs from the Timepix/Medipix chip family.
It features an active area of \SI{3.2 x 3.2}{\milli \meter} housing $128 \times 128$ pixels with a pitch of \SI{25 x 25}{\micro \meter}.
Each pixel is equipped with a 5-bit time-over-threshold (ToT) and 8-bit time-of-arrival (ToA) measurement.

The chip operates in a shutter-based data acquisition mode, adapted to the beam conditions expected at CLIC, and also implements power pulsing of several digital-to-analog converters (DACs) in the pixel matrix.
Since the chip was fabricated as part of a multi-project wafer run, a major challenge to overcome was bump bonding of the ASIC at single-chip level with \SI{25}{\micro \meter} pitch.

First promising results from laboratory and test beam measurements have recently been presented~\cite{clicpix2-pisa}, including a spatial resolution of $\sigma_{SP} \approx \SI{5}{\micro \meter}$ at a sensor thickness of \SI{130}{\micro \meter}.
This indicates that the target resolution for the CLIC vertex detector of \SI{3}{\micro \meter} is not achievable with \SI{50}{\micro \meter} thick standard planar silicon sensors at a pixel pitch of \SI{25}{\micro \meter}.

\subsection{Power Pulsing of Front-End Electronics}

Owing to their beam structure, linear colliders have a very low duty cycle; At CLIC collisions occur during less than \SI{0.01}{\permille} of the time.
The idea to save power by switching some parts of the front-end electronics into a lower-power idle state between bunch trains is therefore attractive and has been studied in the past~\cite{blanchot-dannheim}.

Several ASICs such as Timepix3~\cite{timepix3,timepix-powerpulsing}, CLICpix, CLICpix2 and CLICTD implement power pulsing functionality for different parts of the circuitry, and the power pulsing behavior of the CLICpix2 prototype has been studied in detail as a function of different parameters~\cite{sara-report}.
Here, power pulsing of the analog front-end in the pixel cells is implemented by multiplexing the supply currents to the relevant components between two DACs.
One DAC provides the full current to operate the front-end (high-power state) while the other DAC provides a significantly reduced current for the low-power idle state.
For CLICpix2 only the preamplifier and discriminator can be switched.
An inclusion of further DACs into the power pulsing scheme can be envisaged for future chip versions.

When ramping up the current on the preamplifier and shaper, a so-called \emph{power-on response} from the pixel front-end is expected, and depending on the low-power level it takes several microseconds for the chip to become quiet and ready for data acquisition.
For the default values of the low-power state in CLICpix2, this delay is measured to be \SI{6}{\micro \second}, while reducing the low-power state to zero increases this time to about \SI{55}{\micro \second}.
However, when calculating the average analog power consumption over a full CLIC accelerator cycle, the low-power idle state is the dominating factor and outweighs the time difference in the high-power state.

A total reduction of the analog power consumption by a factor five from \SI{980}{mW/cm \squared} to \SI{190}{mW/cm \squared} has been measured.
With inclusion of additional DACs in the power pulsing, a reduction by a factor 80 can be achieved, leaving the chip with a total analog power consumption of less than \SI{12}{mW/cm \squared}.

\subsection{Hybridization with Anisotropic Conductive Film}

Alternatives to conventional solder-bump bonding of hybrid pixel detector assemblies are studied to overcome the problems related to cost and limitation of the pixel pitch.
One approach of replacing solder balls is hybridization using anisotropic conductive film~\cite{acf}.
Here, conductive micro-particles with a diameter of about \SI{3}{\micro \meter} are randomly distributed over an adhesive film with a thickness of about \SI{18}{\micro \meter}.
On the readout ASIC and sensor, bond pads are formed using an electroless nickel immersion gold (ENIG) process.

By applying force, the adhesive film is squeezed and the micro-particles between the bond pads of the two dies are deformed and establish an electrical connection.
This technique is widely used e.g.\ in the LCD display manufacturing industry along one dimension.
The challenge for its application to silicon pixel detectors is to optimize the parameters such as tacking force, film thickness, micro-particle diameter and quantity for a uniform distribution in two dimensions.
Currently this endeavor is in an early R\&D phase.
Promising first results have been achieved using glass samples as well as ASIC-to-ASIC assemblies for cross-sections allowing visual inspection of particle distribution and bonding, as well as dedicated test structures for yield and resistance measurements.

\section{Monolithic Pixel Detectors}
\label{sec:monolithic}

In contrast to hybrid pixel detectors, monolithic detectors combine sensing element and electronics circuits in the same wafer.
In order to achieve a sizable depletion volume and therefore a fast signal formation, the electronics need to be shielded with deep wells from the higher voltages applied to the sensitive volume.
There are different possibilities to arrange the collection electrode with respect to the electronics.
When the shielding well encloses the pixel front-end electronics completely and also acts as collection electrode, a high voltage can be applied to the wafer substrate leading to a large depletion zone.
However, at the same time the capacitance of the electrode is rather large, which leads to a comparatively high power consumption and lower signal to noise ratio.
Alternatively, a small collection electrode can be placed next to the shielding well, which reduces the input capacitance leading to a low noise and power consumption.
Here, only smaller voltages can be applied and the depleted region is formed by using a high-resistivity epitaxial layer.
Process modifications allow full lateral depletion also in these designs~\cite{tj-modified} by adding an additional deep N-layer.
This allows using higher backside bias due to a better isolation of the electronics by the depleted region.

\subsection{The CLICTD Prototype}

The CLICTD prototype is a fully-integrated sensor designed for the requirements of the CLIC tracking detector~\cite{clicdp-conf-2019-009,clicdp-conf-2018-008}, fabricated in a \SI{180}{\nano\meter} CMOS imaging process with a small N-well collection electrode on a \SI{30}{\micro \meter} thick P-type high-resistivity epitaxial layer.
Two different sensor designs have been produced, one with a uniform N-layer implantation, the other one with additional gaps along one dimension to reduce collection time and in-channel charge sharing.

The chip features a matrix with $16 \times 128$ pixels with a pitch of \SI{300 x 30}{\micro \meter} and a total sensitive area of \SI{4.8 x 3.84}{\milli \meter}.
It is operated in a shutter-based mode suitable for linear colliders.
The analog front-end of each pixel is divided into eight sub-pixels along the \SI{300}{\micro \meter} pitch to ensure prompt charge collection.
The digital front-end provides 8-bit ToA and 5-bit ToT measurements, combining the sub-pixel comparator outputs through a global \emph{OR}, as well as the hit patterns of the sub-pixels.

\subsection{Laboratory Calibration}

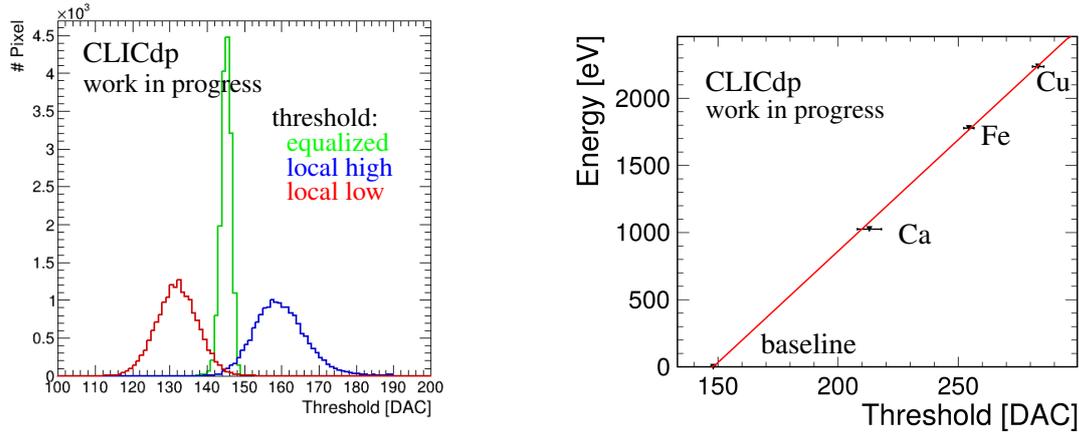
\begin{figure}[tbp]
        \centering
        \begin{subfigure}[t]{0.48\textwidth}
                \begin{overpic}[width=.9\textwidth]{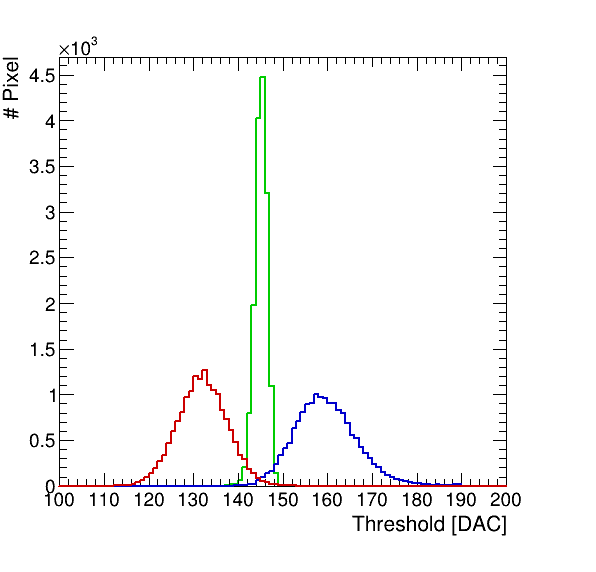}
                        \put (15,78) {CLICdp}
                        \put (15,72) {\small work in progress}
                        \put (53,65) {\small threshold:}
                        \put (56,60) {\small \textcolor{green}{equalized}}
                        \put (56,55) {\small \textcolor{blue}{local high}}
                        \put (56,50) {\small \textcolor{red}{local low}}
                \end{overpic}
        \end{subfigure}
        \begin{subfigure}[t]{0.48\textwidth}
                \begin{overpic}[width=\textwidth]{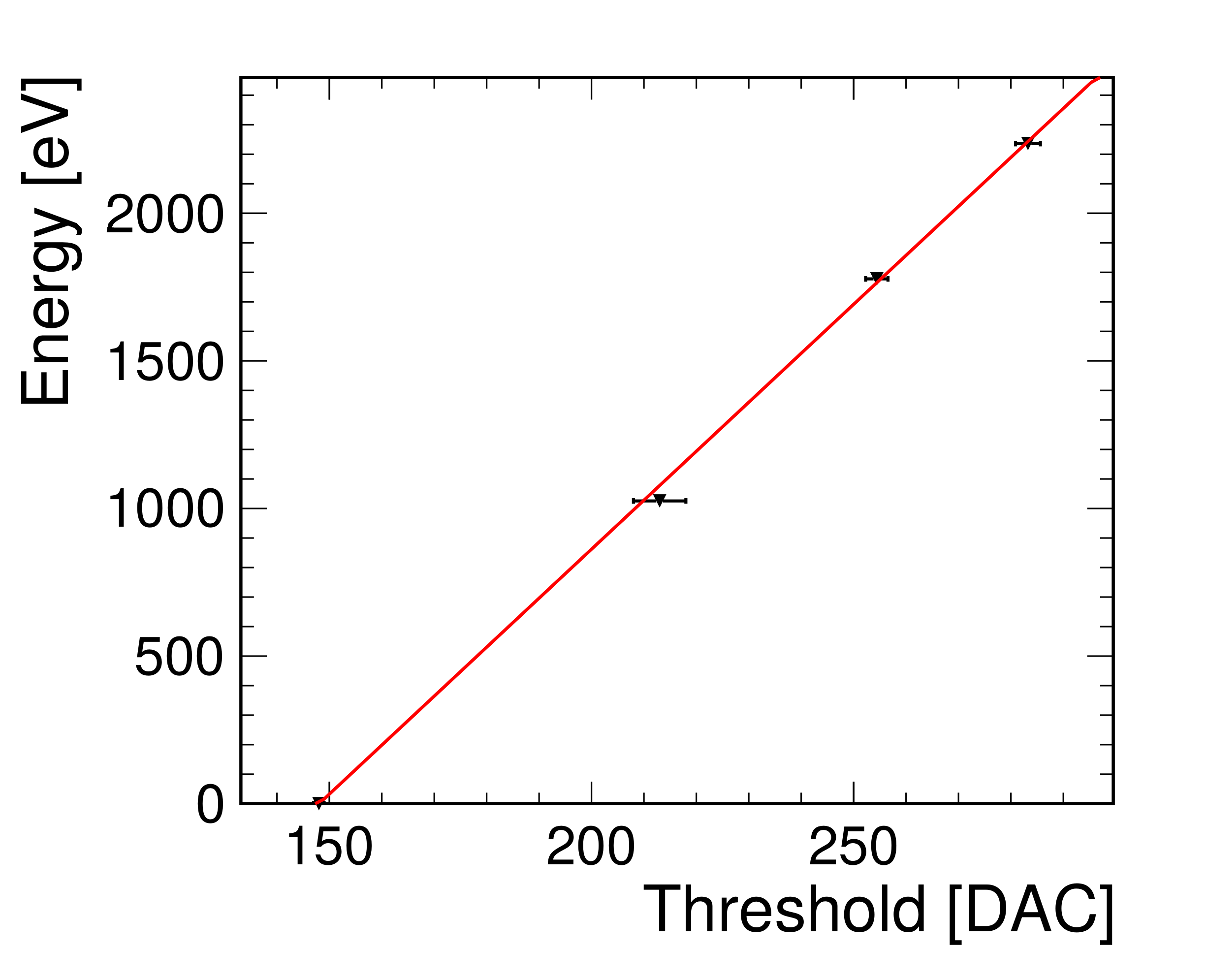}
                        \put (25,65) {CLICdp}
                        \put (25,60) {\small work in progress}
                        \put (85,65) {Cu}
                        \put (75,55) {Fe}
                        \put (60,37) {Ca}
                        \put (35,17) {baseline}
                \end{overpic}
        \end{subfigure}
        \caption{Laboratory calibration of threshold equalization and energy for the CLICTD prototype: Pixel threshold distribution before and after equalization (left). Threshold calibration of CLICTD using florescence X-rays from different target materials (right).}
        \label{fig:clictdcal}
\end{figure}

Initial laboratory tests of the chip have shown that the ASIC can be operated as expected.
It has been observed that the circuitry in the P-wells is sensitive to the applied bias voltages.
This results in a correlation between the circuitry response and the sensor operation parameters, and various parameters of the chip have to be tuned for every setting of the bias voltages.

The per-pixel threshold can be equalized over the matrix by using local adjustments as shown in Figure~\ref{fig:clictdcal}~(left).
The threshold as well as the ToT response of the chip have been calibrated on a pixel-by-pixel basis using the characteristic $K_\alpha$ X-ray lines of different materials such as iron, copper and calcium as shown in Figure~\ref{fig:clictdcal}~(right).
A threshold dispersion of \SI{25}{electons} and a pixel noise RMS of \SI{13}{electons} have been measured, other parameters are being investigated.

\subsection{First Test Beam Results}

\begin{figure}[tbp]
        \centering
        \begin{subfigure}[t]{0.48\textwidth}
                \begin{overpic}[width=\textwidth]{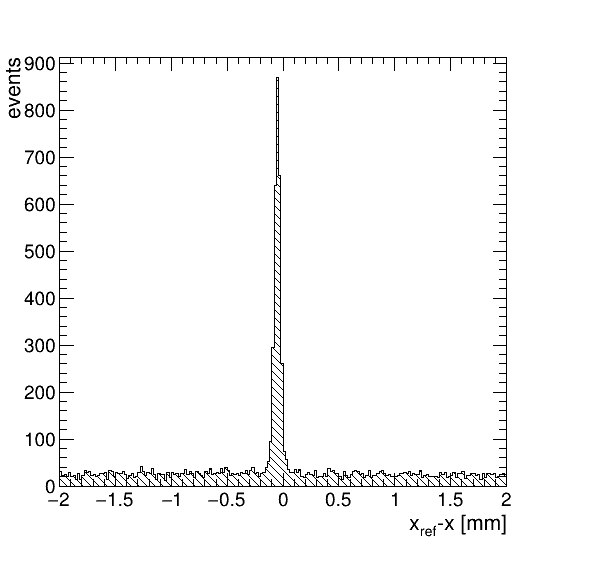}
                        \put (13,77) {CLICdp}
                        \put (13,72) {\small work in progress}
                \end{overpic}
                \caption{Spatial residual telescope - CLICTD}
                \label{fig:clictd:corrx}
        \end{subfigure}
        \begin{subfigure}[t]{0.48\textwidth}
                \begin{overpic}[width=\textwidth]{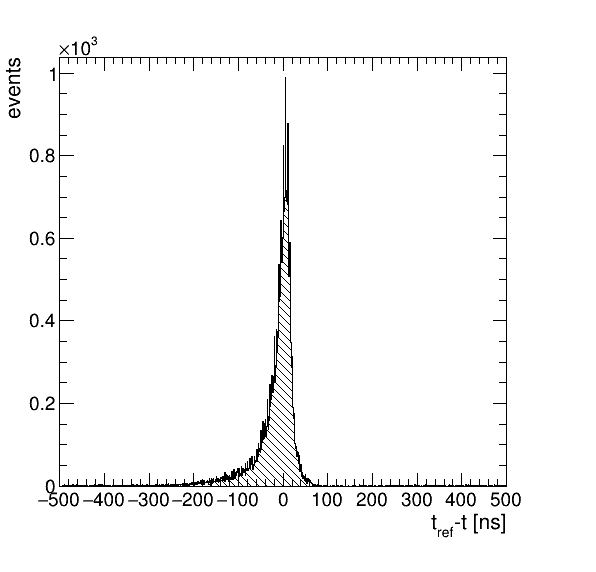}
                        \put (13,77) {CLICdp}
                        \put (13,72) {\small work in progress}
                \end{overpic}
                \caption{Time residual Timepix3 - CLICTD}
                \label{fig:clictd:corrt}
        \end{subfigure}
        \caption{Spatial and time residuals of CLICTD as recorded by the online data quality monitor during test beam measurements.}
        \label{fig:clictd}
\end{figure}

Recently, first test beam campaigns with CLICTD have successfully been conducted at the DESY T21 beam line using a \SI{5.4}{GeV} $e^-$ beam and first residual plots are shown in Figure~\ref{fig:clictd}.
Reference tracks have been recorded using the DATURA telescope~\cite{pub-datura-paper} and additional timing information was provided by a Timepix3 plane placed downstream of the setup.

In CLICTD, the ToA is measured with respect to the closing of the acquisition shutter.
With a random arrival time of the beam particles, the maximum shutter length is limited to \SI{2.5}{\micro \second} by the size of the 8-bit ToA counter.
In order to overcome this limitation, the scintillator trigger indicating the particle arrival has been used to close the shutter of the detector, leading to a fixed distance in time between the two events while allowing to open the shutter already much earlier and thus increasing the data taking efficiency.
% This mode of operation allows to directly access the time resolution of the device.

The full analysis of the data is currently under way and more test beam campaigns are scheduled to investigate the different effects observed in laboratory measurements and to study the detector performance for inclined particle tracks.

\section{Tools for Silicon Detector R\&D}
\label{sec:tools}

Versatile tools are required to qualify different prototypes and to investigate the large variety of silicon detector technologies under consideration for the CLIC vertex and tracking detectors.
Several tools have been developed within the CLICdp collaboration over the past years and will be described briefly in the following sections.
They have proven their flexibility and performance and are used by other groups in the silicon detector R\&D community.

\subsection{Pixel Detector Data Acquisition with Caribou}

Developing a new silicon detector requires significant effort for preparing the readout hardware and software for the prototype to be operated in the laboratory and test beams.
The \emph{Caribou DAQ system}~\cite{caribou} has been developed to significantly reduce the effort and cost involved in designing such a system from scratch for every new chip.
It combines programmable logic and a processing system by utilizing a System-on-Chip (SoC) platform and thereby brings unprecedented flexibility to the DAQ design.
An interface card connects the SoC with the detector prototype, housing power supplies for biasing as well as DACs and ADCs for setting and measuring operational parameters, test pulses, etc.
The system is completed by a set of configurable firmware blocks for commonly used functionality as well as the DAQ software Peary.
The latter is fully integrated into the EUDAQ2 framework~\cite{eudaq2} and no further work is required to operate new prototypes in complex test beam environments.
The FPGA fabric is available for detector control and data handling, while the CPU runs a full Linux distribution providing the possibility to directly interact with the detector via command line tools or the EUDAQ2 producer.
The system already supports a large set of different prototypes, including H35Demo~\cite{h35demo}, ATLASPix1--3~\cite{atlaspix}, CLICTD, CLICpix2 and C3PD~\cite{c3pd}.

\subsection{Test Beam Data Reconstruction with Corryvreckan}

The complexity of test beam data reconstruction lies in the combination of very different detectors, many of them in prototyping stage only, with varying readout schemes.
Especially when combining trigger-based detectors designed for the LHC and frame-based detectors for linear colliders, a flexible offline event building algorithm is required.
For this purpose, the \emph{Corryvreckan} reconstruction and analysis framework has been developed~\cite{corry-repo}.
It is a flexible, highly configurable software with a modular structure designed to reconstruct and analyze both test beam and laboratory data.
Special emphasis has been put on providing a detailed documentation in form of a user manual and in-code class documentation.
It provides full integration of EUDAQ2 and is able to directly read data from any detector recorded in this format.
The framework is capable of using timing information for clustering and tracking when available.
For alignment the Millepede-II algorithm is implemented~\cite{millepede2}.

Corryvreckan has been used extensively in the test beams carried out for the prototypes presented above.
Most notably, it has performed very well when combining data from the data-driven Timepix3, the rolling shutter DATURA telescope and the frame-based CLICTD devices.
For synchronizing the different devices, a common clock and start signal (T0) as well as trigger information can be used.
Support for advanced track models such as General Broken Lines~\cite{gbl} is currently under development.

\subsection{Silicon Detector Monte Carlo Simulation with $\mathbf{Allpix^2}$}

Understanding the detector performance of prototypes and optimizing their design before production requires a Monte-Carlo simulation of the full detection chain, including stochastic effects and fluctuations in the signal formation and digitization as well as the production of secondary particles.
For this purpose, the $\mathrm{Allpix}^2$~\cite{allpix-squared} simulation framework has been developed within the CLICdp collaboration, with the aim to provide a generic simulation platform for silicon detectors which eases the implementation of new simulation models, provides an interface to custom electrostatic field simulations, and which is well documented to lower the barrier for new users.

The framework has seen almost 20 feature and patch releases since its first stable release in summer 2017, and has found wide-spread use within and outside of high-energy particle physics.

\section{Summary \& Outlook}
\label{sec:summary}

The experimental conditions at CLIC, a proposed linear $e^+e^-$ collider, pose unique challenges to silicon detectors such as an excellent spatial and time resolution while maintaining a minimal material budget. A comprehensive R\&D program is underway for silicon detectors at CLIC, encompassing different technologies and detector concepts.
Dedicated prototypes have been developed and qualified in laboratory and test beam measurements.
The measurements have shown that most initial requirements are already achievable with the technology available today, but further effort concerning detector integration is necessary and the spatial resolution of \SI{3}{\micro \meter} at a sensor thickness of \SI{50}{\micro \meter} still has to be reached.
A set of versatile tools for simulation, data acquisition and reconstruction has been developed as part of the R\&D efforts, which are by now widely used in the silicon detector community.

Many developments are still ongoing, such as the qualification of anisotropic conductive film as hybridization method, and test beam campaigns are planned to qualify prototypes.

\bibliographystyle{unsrt}
\bibliography{bibliography}

\end{document}